# Quantum theory is not only about information

*Laura Felline*

Department of Philosophy, University of Rome III

**Abstract**. In his recent book *Bananaworld. Quantum mechanics for primates*, Jeff Bub revives and provides a mature version of his influential information-theoretic interpretation of Quantum Theory (QT). In this paper, I test Bub's conjecture that QT should be interpreted as a theory about information, by examining whether his information-theoretic interpretation has the resources to explain (or explain away) quantum conundrums. The discussion of Bub's theses will also serve to investigate, more in general, whether other approaches succeed in defending the claim that QT is about quantum information. First of all, I argue that Bub's interpretation of QT as a principle theory fails to fully explain quantum non-locality. Secondly, I argue that a constructive interpretation, where the quantum state is interpreted ontically as information, also fails at providing a full explanation of quantum correlations. Finally, while epistemic interpretations might succeed in this respect, I argue that such a success comes at the price of rejecting some in between the most basic scientific standards of physical theories.

**Introduction**

In the last decades, information theory and its applications to the investigation of the quantum world have experienced a great development. Part of the interest around this area of research comes from fact that within information-based approaches, time-honoured conundrums of Quantum Theory (QT) seem to disappear. For instance, the exclusion of any reference to the mechanical elements underlying quantum phenomena (i.e. exclusion of any assumption about what kind of entities quantum systems are, and what kind of processes they undergo in virtue of their properties) seems to allow the explanation away of the measurement problem or of the problem of non-locality. This apparent success, in turn, contributes to the idea that information theory is the best framework for the formulation of QT and that information plays a special role in the foundations of QT. In particular, one of the most discussed articulations of this idea, is that information is, in a sense to be further specified, *the* subject matter of QT, i.e. that QT is about quantum information, and only about information.

The foundational tenability of the claim that QT is about quantum information has already been subject of profound scrutiny both formally and conceptually (e.g. Hagar and Hemmo 2006, Timpson 2013); in this paper, I will focus on the consequences that this claim has at the level of the theory of explanation. I take it as a virtue of the following arguments, that they hold independently of the points already made in the above cited, previous works. The obvious risk of informational approaches to QT is that the rejection of a more committed description of the physical world comes at the expenses of the explanatory power of the theory. If this risk is real – if, that is, there are quantum phenomena or issues that are non-explainable (nor explainable away) within QIT, then, I will argue, such explanations should be searched for within another interpretation of the theory, which means that QT cannot be *uniquely* about information.

In his recent book *Bananaworld. Quantum mechanics for primates*, Jeff Bub revives and provides a mature version of his influential information-theoretic interpretation of quantum theory. In this paper, I test Bub's conjecture that QT is about quantum information by examining whether his information-theoretic interpretation has the resources to explain (or explain away) quantum phenomena, while at the same time overcoming the issues affecting other interpretations of QT.

Since the significance of Bub's proposal in the debate over the foundations of QT can hardly be overrated, an assessment of this new version of his claim that QT is about quantum information is imperative.

A recurrent element in Bub's analysis of QT is the claim that QT should be understood as a theory of principle based on information-theoretic principles, in the same way as special relativity is a theory of principle based on the relativistic principles. Such a thesis was firstly articulated in its most complete form in the context of the characterization theorem formulated by Bub himself together with Robert Clifton and Hans Halvorson (Clifton *et al.* 2003). In his last book, this framework is substituted with axiomatic reconstructions of QT based on PR-boxes (Popescu and Rorlich 1994).

Here's Bub suggestive illustration of his view.

> On the information-theoretic interpretation, the relation between quantum mechanics and the structure of information is analogous to the relation between special relativity and the structure of space-time. Heisenberg's "re-interpretation" of classical quantities as noncommutative, or more specifically the entwinement of commuting and noncommuting observables, imposes objective pre-dynamic probabilistic constraints on correlations between events in a similar sense to how Minkowski space-time imposes kinematic constraints on events. The probabilistic constraints of the correlational structure provide the framework for the physics of a genuinely indeterministic universe. They characterize the structure of information for nonlocal correlations like Popescu-Rohrlich correlations, which can only occur between intrinsically random events. (2016, p. 223)

This also grounds his outline of how QT explains non-local correlations, in that it justifies the claim that information-theoretic interpretations explain quantum correlations in the same way in which special relativity explains length contraction. I will argue that such a suggestion should be understood in the sense that QT provides a kind of explanation of non-local quantum correlations where the explanandum is shown to be an instantiation of a fundamental structure of the world, rather than mechanically produced by some underlying process (Felline 2015).

In the process of explicating Bub's suggestions about an information-theoretic explanation of quantum correlations, I will show that the success of this project depends on the presence of an upstream account of the determinateness of our experiences. Without such an account, no explanation of non-local correlations can be said complete. While I will criticize the possibility that an ontic interpretation of the quantum state provides said account, this discussion provides the occasion to broaden the scope of the proposed analysis beyond Bub's own proposal.

The proposed discussion of Bub's work will therefore serve to investigate, more in general, whether other interpretations succeed in providing a satisfactory account of the determinateness of our experiences, that might work as a basis to the claim that QT is about quantum information. As the arguments presented in the paper will unfold, I will progressively apply them to different variants of the claim that QT is about information, constructive approaches (both ontic and epistemic) included.

The first two sections are devoted to the analysis of the significance of the principle/constructive dichotomy, which will be instrumental for the inquiry put forward in the rest of the paper – not only with respect to Bub's principle approach. It should be right away pointed out that it is not the scope of these two sections to provide an exegesis of this dichotomy that faithfully unfold Einstein's philosophical view. In section 1 I argue that the principle/constructive dichotomy is a distinction about two features of a physical theory: the kind of basic assumptions on which it is built ("basis and starting points") and the kind of inferential arguments that are used to get from the basic assumptions to the rest of the theory (analytic or synthetic method). In the second section I illustrate and criticize some widespread assumptions about the relationship between explanation and the principle/constructive dichotomy (which, again, is not related to the claim that this is what Einstein *really* meant).

This analysis will be instrumental in the next sections. In particular, it will be central in the analysis of Bub's proposal that QT, as a principle theory about information, provides an explanation of quantum correlations that traces the same kind of explanation provided in special relativity of, say, length contraction (section 3). In order to evaluate this claim, I will provide a detailed account of how special relativity explains length contraction and illustrate the parallel explanation of correlations in the information-theoretic approach.

The conclusion of this section is that QT as a principle theory about information fails to successfully explain quantum correlations. The claim that information is fundamental, in fact, can't be derived from information-theoretic principles, so that it can't be considered part of a principle interpretation of QT. On the other hand, the claim that information is fundamental is necessary for an information-theoretic explanation of quantum correlations, and since a principle reconstruction of QIT does not include such a claim, then a successful explanation of quantum correlations requires a constructive interpretation.

The claim that QT is about quantum information cannot therefore be implemented within a principle theory about information-theoretic principles. Section 4 investigates whether this goal can be instead achievable by a constructive QIT. In this context, I show how a successful explanation of quantum correlations requires a pre-existing solution to the measurement problem, and I argue that ontic interpretations fail in this respect.

In section 5 I show how epistemic interpretations of the quantum state formally succeed in explaining away the measurement problem, but in section 6 I critically analyze their solution and put forward a challenge that that they have so far failed to meet.

1. **What the principle-constructive distinction actually says**

Einstein's notorious characterization of the constructive/principle theories distinction appeared in an article for The Times London (Einstein 1919), titled originally "Time, Space, and Gravitation.", then reprinted as "What is the Theory of Relativity?" (Einstein 1954). It is often considered one of Einstein's most insightful and fruitful contribution to philosophy of science, and a lively debate is still ongoing on the real significance and impact of this distinction.

So, here it is.

> We can distinguish various kinds of theories in physics. Most of them are constructive. They attempt to build up a picture of the more complex phenomena out of the materials of a relatively simple formal scheme from which they start out. Thus, the kinetic theory of gases seeks to reduce mechanical, thermal, and diffusional processes to movements of molecules—i.e., to build them up out of the hypothesis of molecular motion. When we say that we have succeeded in understanding a group of natural processes, we invariably mean that a constructive theory has been found which covers the processes in question.
>
> Along with this most important class of theories there exists a second, which I will call "principle-theories." These employ the analytic, not the synthetic, method. The elements which form their basis and starting-point are not hypothetically constructed but empirically discovered ones, general characteristics of natural processes, principles that give rise to mathematically formulated criteria which the separate processes or the theoretical representations of them have to satisfy. Thus the science of thermodynamics seeks by analytical means to deduce necessary conditions, which separate events have to satisfy, from the universally experienced fact that perpetual motion is impossible.
>
> The advantages of the constructive theory are completeness, adaptability, and clearness, those of the principle theory are logical perfection and security of the foundations.
>
> The theory of relativity belongs to the latter class. (Einstein [1919] 1954, p. 228)

As characterized by Einstein, the principle/constructive distinction does not directly concern explanation, ontology or causation. More simply, it concerns two features of a scientific (physical) theory: the kind of basic assumptions on which it is built ("basis and starting points") and the kind of inferential arguments that are used to get from the basic assumptions to the rest of the theory (analytic or synthetic method).

First of all, the basic assumptions of principle theories are general, phenomenological, principles. The starting points of constructive theories are 'hypothetically constructed', and concern the elementary constituents with which the theory builds up a picture of more complex phenomena.

Secondly, the analytic method with which theories of principle are built requires a formalization of the principles and a rigorous mathematical deduction of the rest of the theory as theorems of the axiomatic system. On the other hand, constructive theories are built with a synthetic method, which does not only apply to the formulation of the basic principles (e.g. the hypothesis of molecular motion), but also to the subsequent 'reduction' or reconstruction, of the more complex phenomena. Although, in fact, constructive theories are also mathematized and make essential use of mathematical derivations, yet other kinds of considerations are also relevant for the evaluation of the mathematical inferences in the theory. It is because of their specific methodologies that constructive theories are 'adaptable', while principle theories are 'logically perfect'.

That's all Einstein's characterization strictly says. However, there are other considerations that can be made by starting from such simple features.

For instance, constructive theories are more likely than principle theories to produce claims about what kind of entities and processes constitute the world – or, at least, this is so in the domain of microphysics, where phenomenological principles cannot directly describe what there is in the world. While, in fact, this kind of claims can be more easily included in the starting hypothesis of constructive theories, the only claims that are strictly part of a principle theory are those that can be logically/mathematically derived from the empirical generalizations that act as principles. However, as we will see later in the paper, nothing in Einstein's characterization requires that the hypothetical claims of constructive theories concern the 'stuff' (entities and processes) that constitutes the world, as they can, for instance, concern structural facts.

With respect to the kind of explanations provided, we can say that the availability of a richer set of entities and processes typically provides constructive theories with the material for richer mechanistic explanations of the kind described, for instance, in (Machamer *et al.* 2000). However, this does not straightforwardly imply that causal explanations only occur in constructive theories. First of all, the commitment to what kind of entities and processes underlies a phenomenon is not necessary to causal explanations in the sense of relevance causation[1] (e.g. Woodward 2005). Secondly, some accounts of causal explanation in the sense of productive causations might apply to explanations in principle theories, if an ontological commitment over the kind of entities involved in the production is not relevant (e.g. Salmon-Dowe's conserved quantity account (Salmon 1994) applies to thermodynamics). Third, it might be said that mechanistic explanations are also provided within some principle theories, e.g. the explanation of the behavior of a heat engine within classical thermodynamics.

One typically neglected aspect, but that will play a pivotal role in this paper, concerns the epistemic import of the distinction. The general principles at the basis of principle theories are phenomenological generalizations of empirical data, while the hypotheses at the basis of constructive theories have been formulated as explanatory hypotheses of the more complex phenomena that constitute the domain of the theory. One notable consequence of this feature is that different criteria for scientific evidence and justification usually apply to these theories. For instance, the starting hypotheses of constructive theories must have implications that are independent (and independently testable) of the very same 'complex phenomena' for which the theory is intended to build up a picture. In fact, if the only empirical support available for such starting hypotheses would be that they can explain the phenomena the theory was called to explain in the first place, then the resulting explanation would be *ad hoc*. To stick to Einstein's example, the hypothesis of molecular motion gained a stronger scientific status – and eventually won over the alternative hypothesis of energetism – thanks to the corroboration provided by the fact that it was able to explain the convergence between the various different procedures for the measure of Avogadro's number. Evidential support in principle theories work in a very different way. The same requirement of independent controllability does not apply equally to principles of principle theories, since they are already justified by empirical evidence. For instance, no further justification is necessary for the principles of thermodynamics, or to the principles of relativity, other than the fact that they are highly empirically confirmed.

---

1 For the concepts of relevance and productive causation see (Hall 2004).

2. **Some widespread, but wrong, assumptions about explanation in principle and constructive theories**

The short passage cited above that '[w]hen we say that we have succeeded in understanding a group of natural processes, we invariably mean that a constructive theory has been found which covers the processes in question" has been the subject of an extended literature, and adduced as warrant for a variety of claims about the explanatory features of principle and constructive theories.

However, at the basis of such claims often lies a simplistic idea of the relationship between explanation and the principle/constructive dichotomy. Before passing to the discussion of explanation in QIT it is useful to linger on some widely spread misleading assumptions, that might influence the evaluation of axiomatic reconstructions of QT, and in particular of Bub's account of QT as a principle theory about the structure of information.

A very influential view states that principle theories are explanatorily defective. Harvey Brown and Oliver Pooley (2006) have famously used this claim in their criticism of the 'orthodox' interpretation of special relativity that, they argued, lacks explanatory power. As a general point, it should be said that measuring the explanatory power of a theory is a more complex matter than assumed by this claim. If talking of a 'measure' of the explanatory power of a theory makes any sense at all, then such a measure must take into account, just to name few factors, what kind of questions emerge in the domain of the theory, the kind of explations that one is expected to find (causal, unificationist, counterfactual) and the theoretical tools with which the theory approaches its domain. The principle and constructive distinction does not exhaust such questions.

Let's therefore consider some specific arguments to the conclusion that principle theories are explanatorily weak.

Brown and Pooley have argued that explanations in principle theories are deductive-nomological explanations. If this would be true, it might be argued, then principle theories explanations would in fact be clearly weak, as mere deductive arguments starting from the relativistic principles.

Against Brown and Pooley, however, other works better characterize the explanatory richness of specific principle theories, and all of them provide a stronger sense of explanation than the deductive-nomological model. Flores (1999) underlines the explanatory power of principle theories in terms of their unifying power, while Van Camp (2011) elaborates on the constitutive conceptual role played by space-time theories (DiSalle 2006) and argues that "[i]n some cases, the primary function of principle theories is to establish principles which are constitutive of the very framework of some set of physical concepts. […] theories such as Newton's and Einstein's play a fundamental explanatory role by establishing the explanatory framework itself." (p.7). Felline (2016) applies a counterfactual account of explanation to explanations in axiomatic reconstructions of QT in terms of information-theoretic principles.

None of these explanatory virtues can be captured by the claim that principle theories provide mere deductive-nomological explanations.

Another influential characterization of explanation in principle and constructive theories was articulated by Balashov and Janssen (2003). They agree with Brown and Pooley that principle theories are associated with deductive-nomological explanations,[2] but they add that constructive theories, and only them, provide model-based explanations that unveil the 'reality behind the phenomena'. A first comment on this characterization is that it is false that principle theories have nothing to say about the 'reality behind the phenomenon'. Thermodynamics, for instance, tells us that isentropic processes are always adiabatic processes. Secondly, it is not true that only constructive theories provide model-based explanations. Any scientific theory, theory of principles included, provide models and such models can be used to provide model-based explanations. After all, any explanation can be accounted for as model-based explanation, deductive-nomological explanations included. It is sufficient: i) to provide a model of the relevant domain of the explanandum phenomenon, that is true of the laws and initial conditions that should be used by the deductive-nomological explanation; ii) to formulate an explanation within the model, by subsuming the representative of the explanandum under the model's laws of nature (Felline 2011).

It also has to be said that, while it is true that some phenomena are explainable by constructive theories but not within principle theories, the opposite is also true. For instance, traditional (constructive) interpretations of QT revealed so far ineffective in providing an answer to the question 'why is the quantum world non-local?'. In fact, some philosophers claim that this is a structural limitation of interpretations of QT (see e.g. Egg and Esfeld 2014). On the contrary, principle reconstructions of QT in terms of information-theoretic principles can address questions like 'why is the quantum world non-local?', or 'why is our world only this much non-local?' (Felline 2016).[3]

Before closing this section, I must address a possible objection. Many of the above formulated objections against claims about explanation in principle and constructive theories, hinge on counterexamples. One, therefore, might contend that every theory possesses both principle and constructive features (Brown and Timpson, for instance, argue that "all theories have principles, it is just that some are more phenomenological than others"[4] (2006, p.5)) and that this non-categorical character of the distinction undermines the arguments and counterexamples illustrated so far. If the distinction between principle and constructive theories is not categorical, in fact, the examples illustrated above can be interpreted as reflecting the double nature (principle and constructive) of scientific theories, rather than contradicting the claims about explanation in principle and constructive theories. For instance, the fact that thermodynamics shows that all processes are adiabatic can be interpreted as showing the double principle/constructive nature of thermodynamics, rather than the fact that also principle theories unveil the reality behind phenomena.

---

[2] It should be said that, in Janssen's analysis, the explanatory power of special relativity does not hinge heavily on its being a principle or a constructive theory, but rather on the kinematic/dynamics distinction. It should also be said that Bub has been recently reconsidering the philosophical significance of the principle/constructive distinction, towards a reading of the parallel with relativity in terms of the kinematic/dynamics distinction.

[3] In section 3 we will illustrate such explanations more in details.

[4] As a side note: contrarily to what might be suggested by this quote, not any principle can act as basis for a genuine principle theory – only phenomenological, empirical generalization will do. To be sure, the notion of 'phenomenological' might be vague, but this does not mean that any principle can be honestly be said phenomenological. The assumption that the atom is structured as a planetary system, for instance, is clearly not a phenomenological principle.

A natural response to this objection is that the latter is based on a convoluted semantic for the terms 'principle' and 'constructive', which does not match with the simplicity of Einstein's characterization. Although I think that this answer is correct, I also think that, in absence of other considerations, it might wrongly suggest that this controversy rests on purely terminological grounds.

This is far from being true. As I am about to argue, the price for accounting for the explanatory richness of a theory in terms of its mixed principle/constructive character is, in fact, the loss of the real epistemic significance of the distinction.

We have seen above (section 1) that the principle/constructive dichotomy concerns, between other things, the epistemic grounds of theories: it tells us about the kind of assumptions that lie at the foundations of the theory, the logical relationship between basic assumptions and the rest of the theory and it characterizes the evidence required to support the theory.

Let's consider again the claim that only constructive theories unveil the reality behind phenomena, and my counterexample to it, i.e. the fact that thermodynamics says that all process are adiabatic processes. As we have already seen, one could reply to this counterexample by claiming that thermodynamics, as any other theory, is partially constructive, and that this constructive side is responsible for the fact that also thermodynamics unveils the reality behind phenomena. Accordingly, explanations that appeal to the fact that all processes are adiabatic processes should be interpreted as constructive theory explanations.

But here we find a problem. For what we have said before, constructive theories explanations rest on hypothetically construed claims, and the justification for such claims is abductive in nature. But the claim that all processes are adiabatic processes does not possess such a status, being rather a 'theorem' that follows mathematically from the empirical generalizations acting as axioms of thermodynamics as a principle theory.

Therefore, the consequence of taking this argumentative road is the conflict with what I think is one of the most basic and analytically most fruitful uses of the principle/constructive distinction, i.e. distinguishing the epistemic foundations of physical theories (and of the explanations they provide).

The conclusion of this section is that a principle theory does not necessarily lack explanatory power: it does not only provide deductive-nomological explanation, nor it necessarily remains outside the domain of the reality behind the phenomena; it can provide model explanations and it can explain phenomena that constructive theories fail to explain. We must resist an analysis of the explanatory power of QT that merely rests upon its interpretation as a principle theory: such an interpretation does not straightforwardly imply anything specific about the kind, or the success of QT's explanations.

In the next section, we are going to use the distinction with the new understanding of its import stressed in this section.

3. **QT as a principle theory about information explains something, but it does not explain everything**

In this section, we are going to analyse in detail how QT explains phenomena, in the information-theoretic approach.

Bub articulates his new version of the information-theoretic approach with the formal background provided by Popescu's and Rorlich's PR-boxes. PR-boxes are toy-models built from two axioms: relativistic causality, corresponding to no-signalling, and non-locality, neutrally defined in terms of non-local correlations in the sense of Bell's theorem. Popescu and Rorlich carry an investigation based on the manipulation of toy models defined by these two principles and show that quantum correlations are not the sole non-local correlations that are consistent with such a setup. In other words, quantum non-locality is not the sole kind of non-locality allowed in a world where the two principles 'non-locality' and 'relativistic causality' hold, but other, so-called 'post-quantum', correlations exist that can be more non-local than quantum correlations.[5]

This discovery has opened the door to a series of questions about non-locality: if post-quantum correlations don't violate the 'no signalling' principle, why doesn't our world instantiate them? Why is our world only *this much* non-local, when it seems that it could be more? In search for an answer to these questions, further researches (e.g. Brassard *et al.* 2006) have explored the conjecture that the explanation of the existent limit to non-locality lies in a new information-theoretic axiom about the impossibility of trivial communication complexity. In (Pawłowski *et al.* 2009) a new principle is put forward, called information causality, which recites:

> The information gain that Bob can reach about a previously unknown to him data set of Alice, by using all his local resources and m classical bits communicated by Alice, is at most m bits (Pawłowski *et al.* 2009, p. 1101).[6]

Felline (2016), argues that axiomatic reconstructions of QT provide a concrete and quite commonsensical explanation, which explains a feature of the quantum world by showing how this feature depends on what might pre-theoretically be called 'nature' of the quantum world. In this account, however, the notion of 'nature' is replaced with that of *characterizing property*, i.e. a "property unique to a given entity or structure within a family or domain of such entities or structures" (Steiner 1978[7], p.143). The axioms of information theoretic reconstructions of QT are therefore the characterizing property of QT. For instance, the principles of PR-boxes isolate QT against the family of generalized probabilistic theories.

---

5 Here non-locality is measured as the amount of violation of Bell's inequalities.
6 For a short but clear illustration of the project of principle reconstruction of QT in terms of non-local PR-boxes, and for a thorough discussion of the status of the information causality principle see Michael Cuffaro's paper in this volume.

Accordingly, the explanation of quantum non-locality consists in showing how the latter depends on the characterizing property of QT. This implies not only showing that quantum non-locality is entailed by the principles but also how a change in the principles corresponds to a change in the explanandum. This explanation provides therefore a specific kind of 'what-if-things-had-been-different' knowledge, where the counterfactual claims are produced by changing the theory's principles and exploring the consequences of such a deformation for non-locality.

While this explanation is a successful explanation to the question 'why there is non-locality, and only this much non-locality, in the world?', Felline also shows that axiomatic reconstructions of QT fail to address all the aspects of quantum non-locality that call for an explanation. In particular, this explanation fails to dissolve the tension between the natural assumption that correlations are produced by a causal process and the fact that such a process seems to violate Lorentz invariance. In order to solve this tension (which, according to many, represents the main motivation for the quest of an explanation of quantum correlations) it is necessary to provide a description of *how* correlations take place.

Traditional interpretations of QT (with the exception of the *sui generis* Everettian approaches) propose a solution to the problem that describes the details of such underlying processes and attempts to show that their Lorentz invariance is only apparent. Clearly, Bub's interpretation cannot provide this kind of explanation, because information-theoretic constraints are, for their very own constitution, uninformative about such kinds of processes. As we have already anticipated, according to Bub, correlations are explained in QT in the same way as relativistic effects are explained in special relativity, but an evaluation of this proposal requires unpacking. Let's therefore first see in detail how explanations work in special relativity.

In what is probably the most widespread approach to special relativity (the one that Harvey Brown calls the 'orthodox' approach, and the one that Bub himself seems to favour), relativistic effects like length contraction and time dilation are explained as kinematic effects – manifestations of the fundamental geometric structure of the world. In this explanation, the claim that the geometry of spacetime is fundamental plays an essential role, as without such an assumption there would always be scope for (and, indeed, necessity of) a more fundamental theory explaining where the structure of space-time comes from. This, on the other hand, would amount to a dynamical interpretation of special relativity, which is denied in the orthodox interpretation. Following R.I.G. Hughes (1989) we call explanations of this kind, where the explanandum is explained as the manifestation of a fundamental structure, *structural explanations*.

Now, the assumption that spacetime is fundamental does not logically follow from the relativistic principles. Although there are many arguments that strengthen the case for the fundamentality of spacetime (e.g. Norton (2006) and Janssen (2013)), yet this claim does not possess the epistemic status of a theorem of special relativity as a principle theory, but must be hypothetically assumed.

---

[7] This account of explanation in axiomatic reconstructions is heavily inspired by Steiner's (1978) account of explanation in mathematics.

Given the hypothetical status of their basic assumption, it follows that such *structural explanations are not strictly part of special relativity as a principle theory, but they rather belong to a constructive interpretation of the theory itself*.

The same conclusion can be translated in terms of a hypothetical structural explanation of quantum correlations in QIT.

Let's therefore say that non-local quantum correlations should be explained structurally as an instantiation of the structure of information that is not Shannon-like, but von Neumann-like. In order for the structure of information to provide a successful structural explanation of non-local quantum correlations, it is necessary to go beyond the minimal phenomenological interpretation of information, and assume that such a structure is fundamental in the sense that, as the geometry of spacetime, it is not explainable with, or inferable from, the dynamical or constitutive details of underlying particles or waves. On the contrary, if we deny that the structure of information is a fundamental structure, then there must be a more fundamental story that infers/explains the structure of correlations, and we are back to square one: if there is such a story, then the problem of accounting for it with local dynamics appears again and we therefore still need an answer to the question 'how quantum correlations come about?'.

Now, as in the case of special relativity, also here the assumption of the fundamentality of spacetime is hypothetical in nature, since it does not logically/mathematically follow from any of the principles of information-theoretic reconstructions. Such assumption, therefore, is not strictly part of the principle interpretation of QT, but belongs to a constructive interpretation of the theory.

The conclusion of this section is that an information-theoretic interpretation of QT as a principle theory cannot explain quantum correlations. A constructive interpretation of QT as a theory about information is also required.

4. **The ontic information-theoretic interpretation of QT as a constructive theory does not explain quantum correlations**

The conclusion just stated might not represent a serious problem, as, in this scenario, QT might still be in great company: the structural explanation of length contraction is also not provided by special relativity as a principle theory, but it is provided by its constructive, so-called orthodox, interpretation. If Bub's parallel holds, and although the principle interpretation is not sufficient in order to explain quantum correlations, then QT can be interpreted as a theory about the structure of quantum information, in the same way as special relativity is about the structure of space-time.

The devil's in the details, though. A more accurate analysis shows that the parallel between such explanations in special relativity as a theory of spacetime and QT as a theory about information, caves in in some important respects.

First of all, there is a difference between the epistemic status of the hypothesis at the basis of Bub's information-theoretic structural explanation of correlations (i.e. that information is fundamental) and the epistemic status of the hypothesis at the basis of the orthodox structural explanation of length contraction (i.e. that spacetime is fundamental). As already noticed, the hypothesis of the fundamentality of spacetime has an evidential support that is independent of its capacity to explain length contraction, and seems to be justified, for instance, by its pivoting explanatory role also with respect to other facts, as for instance the universality of relativistic phenomena (see Janssen 2013).

The parallel situation in QIT is very different, as the claim that the structure of information is fundamental does not have the same corroboration. Clearly, the conjecture requires additional support other than its supposed capacity of explaining correlations: in lack of other elements corroborating the explanatory conjecture, the resulting explanation would be *ad hoc*.[9]

But this, again, is not necessarily an unsurmountable problem for Bub's account. The fact that some work is needed in order to corroborate the claim that information is fundamental, certainly does not invalidate the information-theoretic account. A more serious problem, though, can be found on further investigation of an information-theoretic explanation of non-local correlations, when such explanation is claimed to be provided by QT as a theory about information.

Let's get back to special relativity. Let's take the explanation of why a specific rod, with proper length $r$ in the inertial frame S, measures $r'$ in S'. The geometrical properties of Minkowskian spacetime, the relative speed between S and S', and the fact that the proper length of the rod is $r$, are all elements of the *explanans*. Clearly, every theoretical investigation must start from somewhere, therefore the uncritical assumption of some facts in the *explanans* is not in itself a drawback. However, if the *explanans* contains a 'problematic' element, in the sense that it, in turn, calls for an explanation (for instance because we have reasons to expect the opposite) then basic scientific spirit dictates that further investigation is necessary in order to achieve a genuine understanding of the group of natural phenomena involved.

In the case of the explanation of the contraction of a specific rod, some elements of the *explanans* are not in turn explainable within special relativity – for instance, the fact that the proper length of the rod is $r$. This, however, does not constitute a flaw in the explanation provided by SR: such a fact does not fall within the domain of special relativity, but it is instead perfectly explainable within another theory. Remember, in fact, that we are here assuming the 'orthodox' kinematic interpretation of special relativity, according to which said theory is independent of any assumption about matter inhabiting spacetime. The explanation of why the proper length of the rod is $r$, instead, requires the description of the constitution of the body itself, and possibly a (causal) history of the construction of the rod, and therefore by assumption the appeal to another theory. In such a theory of matter, on the other hand, the length of a body is not a problematic unexplainable element.

---

8 According to Bub his information-theoretic interpretation of QT can also explain away the measurement problem. However, as we are about to see in the next section, this explanation fails in the ontic interpretation adopted by Bub.

9 But see the last section for more hopeful considerations about this conjecture.

From this point of view, we can therefore say that the explanation provided by special relativity is genuine and successful. In the case of the information-theoretic explanation of non-local correlations we have a similar situation, except that in this case things get problematic.

Let's take an EPR-Bohm experiment and the explanation of why, after Bob's measurement got (say) spin-down, Alice's measurement got spin-up.

The explanation of a specific instance of non-local correlations (i.e. why Alice's measurement result is spin-up when Bob registered spin-down?) in an information-theoretic approach involves the noncommutative algebraic structure of information and the reference to the fact that Bob's measurement yielded spin-down.

We want to make sure that there are no problematic elements hidden in this explanation, in the same way as we have verified that the proper length of a specific rod does not constitute a problematic element in the explanation of length contraction. In particular, the fact that Bob's measurement yielded spin-down should be taken in serious consideration, as implying the determinateness of a measurement result, and therefore the measurement problem. An explanation that appeals to such determinateness by keeping nested the measurement problem is clearly unsatisfactory. This explanation, in fact, does not merely contain the assertion of an unexplained phenomenon, but of a phenomenon that is in desperate need of an explanation! The tension represented by the measurement problem needs to be solved in order to legitimately consider non-local correlations fully understood.

Two scenarios arise at this point.

In the first scenario, as in the case of the kinematic interpretation of special relativity, the outcome of the measurement on a quantum system does not fall within the domain of an information-theoretic interpretation of QT. In this case, the explanation of non-local quantum correlations requires a further theory to be completed, and such a theory must necessarily be QT, whatever this theory might be about: where else finding an explanation (or explanation away) of the result of a measurement in an EPR-Bohm experiment? In this scenario, therefore, a phenomenon belonging to the domain of QT is not in the domain of an information-theoretic interpretation of QT. It follows that the information-theoretic interpretation of QT does not constitute a full interpretation of QT.

In the second scenario, the explanation of correlations can be provided in full information-theoretic terms, therefore an information-theoretic interpretation of QT can also explain why Bob got spin-down. In this scenario, the information-theoretic interpretation of QT provides a full interpretation of QT – but this, again, requires a solution to the measurement problem.

To recapitulate: a structural explanation of quantum correlations calls for a solution to the measurement problem. If the latter is not explainable (or explainable away) within the information-theoretic interpretation, then QT cannot be exclusively about the structure of information.

In his last book, Bub takes the bull by the horns and argues that the information-theoretic interpretation of QT has the resources to solve the measurement problem. In arguing for this claim, he draws from Pitowsky's (2006) distinction between two measurement problems: the 'big measurement problem', which is "the problem of explaining how a measurement produces a definite outcome dynamically" (Bub 2016, p. 223), and the 'small measurement problem', which is the problem of explaining "how a classical probability distribution over macroscopic measurement outcomes emerges dynamically in a measurement process." (*ibid*.)

Let's here skip the issue of how Bub's account deal with the small measurement problem, and let's focus instead on the the big measurement problem, i.e. explanation of the determinateness of measurement results.

According to Bub,

> on the information-theoretic interpretation, the "big" measurement problem is a pseudo-problem. If the universe is genuinely indeterministic and measurement outcomes are intrinsically random because the observable measure is indefinite, then it isn't possible to provide a dynamical explanation of how a system produces a definite outcome when it's measured – that's what it means for the measurement outcomes to be intrinsically random. The random selection of a definite outcome in a quantum measurement process is a feature of the nonclassical structure of information in a quantum world, just as Lorentz contraction is a feature of the Minkowski structure of space-time in a relativistic world. (2016, p. 223)

However, it is false that in a genuinely indeterministic universe 'it isn't possible to provide a dynamical explanation of how a system produces a definite outcome when it's measured'. The theory formulated by Ghirardi, Rimini and Weber (1986), for instance, describes a genuinely indeterministic process that explains the determinateness of our experience – explanation that implies the replacement of Schrodinger's deterministic dynamics with new stochastic dynamics. Since the collapse process is, in this account, genuinely stochastic, there is no way, within GRW, to infer that a measurement yields *that specific* outcome. Inasmuch as inference is a precondition for explanation, then GRW does not explain, but rather explains away, the specific outcome of a measurement, in the sense that it shows that no such explanation is possible, given that the selection is random. On the other hand, GRW is perfectly capable of inferring (and explaining) the fact that a system produces a *definite* outcome when it's measured. *Pace* Bub, the big measurement problem can and should be explained within a theory that describes stochastic dynamics.

The above quote by Bub seems to conflate between two issues. The first consists in the question: 'why the measurement produces *that specific* outcome dynamically?'. This question does not require an answer other than: 'no reason: the selection was random'. The second consists in the question: 'why the measurement produces *a definite* outcome dynamically?', which is neither explained or explained away by the mere stochasticity of the processes that lead to such an outcome, but requires their description.

5. **From ontic QIT to epistemic QIT**

As already mentioned, the distinction between the big and the small measurement problem was originally drawn by Itamar Pitowsky (2006).

The significant difference between Bub's and Pitowsky's approaches is that, while Bub's information-theoretic approach belongs to the ontic interpretations of QT:

> The information-theoretic interpretation is about the structure of probabilistic correlations. It's not about quantum states representing partial knowledge about some ontic state or underlying reality. Rather, the information-theoretic interpretation takes the quantum state as a complete description of a quantum system. (2016, p. 222),

Pitowsky adopts an epistemic interpretation of the quantum state, which "is not a physical object" but rather "a representation of our state of knowledge, or belief" (2006, p. 3).

The same view is again adopted in a joint paper with Bub (which has since changed his position about this):

> On the information-theoretic interpretation, the quantum state is a credence function, a bookkeeping device for keeping track of probabilities – the universe's objective chances" (ibid., p. 8) and "the quantum state is a derived entity, a credence function that assigns probabilities to events in alternative Boolean algebras associated with the outcomes of alternative measurement outcomes. (2010, p.15)

And again:

> we take the quantum state, pure or mixed, to represent a credence function: the credence function of a rational agent […] who is updating probabilities on the basis of events that occur in the emergent Boolean algebra. (*ibid*. p. 19)

Thanks to this feature, Pitowsky's approach does not stumble upon the same mistake as Bub's ontic approach.

Pitowsky rightfully argues that the big measurement problem is 'illusory' in the context of an epistemic interpretation of QT, as it exclusively "concerns those who believe that the quantum state is a real physical state which obeys Schrodinger's equation in all circumstances" (2006, p. 32). If the quantum state only represents our epistemic state, a bookkeeping device for keeping track of probabilities, then the question 'why do we have determinate results' does not require an answer, since the determinateness of our experiences is a fundamental assumption of this interpretation, which means that there is no explanation for it in this context.

The lack of an explanation here is OK because, on the one hand, an explanation must always start from somewhere; on the other hand, such a fundamental assumption is not problematic in this context: why should we expect, in this picture, anything else than determinate experiences? This interpretation does not lead to the measurement problem, because superposed states are just a formal tool, useful for calculating and predicting future phenomena: they do not represent a weird indeterminate state of the outside world, nor of our brain and therefore neither of our mind.

So, Pitowsky is right in saying that, in his interpretation, the measurement problem is 'illusory'. However, as I have shown in the previous section, the stochastic Hilbert space of events does not contribute to the solution to the measurement problem – it is the epistemic interpretation of the quantum state that does all the work in this solution, by putting the determinateness of our experiences as a fundamental fact of the interpretation.

As already mentioned, Pitowsky makes clear that in order to get the measurement problem you need to assume that "the quantum state is a real physical state which obeys Schrodinger's equation in all circumstances". And, as already mentioned, Bub has recently returned to his original ontic interpretation of the quantum state (already present, for instance, in (Clifton *et al.* 2003) and (Bub 2004)), where the quantum state is a real physical state, therefore, a solution to the measurement problem requires a modification or a limitation of the application of Schrodinger's equation.[10] But such a modification implies a dynamical solution to the problem, which Bub also denies.

To sum up: a full information-theoretic interpretation of QT (and a full information-theoretic explanation of non-local quantum correlations) requires a solution to the measurement problem. In this section, we have compared the ontic and the epistemic approach to the measurement problem and argued that the epistemic, but not the ontic, interpretation of the quantum state allows the formulation of a solution to the problem.

6. **A challenge to epistemic interpretations**

Epistemic interpretations of the quantum state are in general able to explain away the measurement problem in the same way as with Pitowsky's approach, explained above. The same solution, for instance, is provided within QBism, according to which "quantum mechanics is a tool anyone can use to evaluate, on the basis of one's past experience, one's probabilistic expectations for one's subsequent experience" (Fuchs *et al.* 2014, p.1).[11]

In *Bananaworld*, Bub motivates his rejection of the epistemic view with the consequences of the theorem formulated in (Pusey *et al.* 2012). Here I focus on a different problem, articulated in terms of a challenge to epistemic approaches, that, I argue, must be met if they want to represent an acceptable solution to quantum conundrums.

As already shown in the previous section, epistemic interpretations explain away the measurement problem by putting the determinateness of our experiences as a fundamental but unproblematic brute fact. They manage to do that by denying that the quantum state represents the outside world, acknowledging in such a way a more or less sophisticated instrumentalist position. With this respect, it has been argued (Timpson 2013 Ch. 9.2) that the position adopted by QBism should be taken separate from standard instrumentalism:

> the quantum Bayesian begins by adopting a form of instrumentalism about the quantum state, but that is far from adopting instrumentalism about quantum mechanics tout court. [the antirealist reading of the quantum state is] conceded at the outset of the programme. And it is conceded in order to serve realist ends. The non-realist view of the state is not the end point of the proposal, closing off further conceptual or philosophical enquiry about the

---

10 Or submit to an Everettian picture, but he also rejects that option.
11 Although in this section I will often make use of QBism as an illustrating example, I believe that the same arguments are applicable to all interpretations of QT that solve the measurement problem through the claim that the quantum state represents our mental state, objective interpretations included (see the end of this section).

> nature of the world or the nature of quantum mechanics; rather it is the starting
> point. (Timpson 2013 p. 209)

Now, if, as Timpson argues, this position does not collapse, at least in its spirit, into standard instrumentalism, it must come with solid motivations for proposing that, for its own nature, a specific physical domain can't be represented, and that the best we can do, when dealing with such a domain, is to represent our mental state. Ideally, such motivations should show that proposals based on the assumption that a realist interpretation of QT is indeed possible (see e.g. Allori 2013), are all doomed to fail.

In other words, we need stringent reasons to acknowledge that quantum phenomena represent the limit to our representative knowledge of the world and that a solution to the measurement problem, based on the description of the measurement process, is not possible.

In his thorough corpus, Bub elaborates an argument that provides just the required rationale for rejecting a mechanical solution to the measurement problem. Although such an argument was formulated in the context of the CBH theorem, as we will see, its relevance extends well beyond that specific version of the information-theoretic interpretation. In his comparison between the information-theoretic approach and traditional mechanical quantum theories, like Bohmian mechanics, Bub has often argued that the explanatory power of traditional interpretations of QT is more apparent than real, with an analysis grounded on a 'methodological principle' that is probably best stated in his (2003):

> if T' and T'' are empirically equivalent extensions of a theory T; and if T entails that, in principle, there could not be evidence favoring one of the rival extensions T' or T'', then it is not rational to believe either T' or T''. (p. 260)

> Now let T be a quantum theory, and let T', T'', … be various extensions of this quantum theory (e.g., Bohm, Everett, etc.). If we accept T; then (by the CBH theorem) we accept that there could be no evidence favoring any one of the theories T', T'' as a matter of physical law. In other words, we accept that there is no possible world satisfying the information-theoretic constraints in which there is evidence favoring one of these extensions over its rivals. (p.261)

So, the methodological moral he draws is that

> a mechanical theory that purports to solve the measurement problem is not acceptable if it can be shown that, in principle, the theory can have no excess empirical content over a quantum theory. By the CBH theorem, given the information-theoretic constraints any extension of a quantum theory, like Bohmian mechanics, must be empirically equivalent to a quantum theory, so no such theory can be acceptable as a deeper mechanical explanation of why quantum phenomena are subject to the information-theoretic constraints. (ibid.)

With the abandonment of the CBH theorem as the basis for the axiomatic reconstruction of QT, the argument above becomes inapplicable; however, in (Koberinski and Müller 2017) a similar argument is put forward to argue that epistemic approaches are more genuinely explanatory than interpretations of QT like Bohmian mechanics or Everettian interpretations. More exactly, at the basis of their argument, is the 'doctrine'

> that concrete, specific claims about the real world should ultimately be backed up by successful empirical predictions which do not also arise identically from rival theories. (Koberinski and Müller 2017, p.5)

Since different claims, at the basis of alternative (e.g. Bohmian or Everettian) explanations of the determinateness of our experiences, provide the same observable predictions,[12] therefore these explanations should be rejected.[13] Their conclusion is that epistemic interpretations of the quantum state are more explanatory than the other (mechanical) constructive interpretation of QT.

Koberinski's and Müller's argument (as well as Bub's) have the virtue of making explicit what I think is a widely-shared motivation for rejecting traditional interpretations of QT.

However, here begins what I think might be a problem for epistemic accounts. It might be countered that, under Koberinski's and Müller's doctrine cited above, epistemic accounts stumble upon the same step as other constructive interpretations, like Bohm's theory. As for the latter, in fact, also the former's explanations (or explanations away, in the case of the measurement problem) rest on a hypothetic claim: that the quantum state does not represent, as in any other physical theory, the state of the part of the world that is in the domain of that theory, but rather represents our state of knowledge or belief. Clearly, this claim does not make any novel empirical prediction, therefore it does not seem to pass Koberinski's and Muller's criterion.

Now, Koberinski's and Müller's doctrine prevents such an objection, given that it specifically applies to claims *about the real world*, while the claim that the quantum state represents our knowledge is not a claim about the real world. On the other hand, if this is what saves epistemic interpretations, then this discrimination between epistemic and ontic theories exposes the fact that, what might be seen as an upside of epistemic interpretations, brings, in this context, to a pyrrhic victory.

According to the distinction just drawn, in fact, while ontic interpretations must submit to the typical epistemic and methodological standards proper of natural sciences (e.g. competing theories should have different predictions), the epistemic view escapes such standards because it is not concerned with the physical world. In other words, epistemic interpretations of QT play a different game, with different rules, with respect to ontic interpretations and, more in general, with respect to natural sciences.

To be clear, there is no logical contradiction in this strategy. However, since these theories play two different games, a fair ratification of this distance is in order. In particular, it is necessary to be adamant in acknowledging that the success of the epistemic over the ontic approach, is due to the fact that the former does not submit to basic standards that constrain the acceptability of theories in natural science – standards like testability, salience, or intersubjectivity (the fact that QBism allows conflicting associations of quantum states by different observers is pivotal in its explanation away of non-locality).

---

12 But see (Hagar and Hemmo 2006).
13 As an anonymous referee notices, Koberinski's and Muller's criterion seems to assume that physical theories are uniquely determined by data – assumption which seems to go against a widely shared view of scientific underdetermination, as formulated most famously by Duhem.

The challenge that epistemic accounts should face, therefore, consists in showing that the same standards constrain and guide theory discovery and choice, rather than being applied arbitrarily if and when convenient. This, in fact, might lead to pernicious consequence in the development of QT.

An especially straightforward consequence of what just said, is the risk that rejecting even part of such standards would lead to the loss of important elements that act as guides to theory choice and discovery and that make scientific inquiry so powerful, or to the demotion of the often praised 'special' epistemic status of scientific knowledge

As an example, take the constraint of intersubjectivity. Obviously, this constraint lies at the foundations of the purported 'objectivity' of scientific knowledge, but let's leave aside this role here. Thanks to the assumption that different observers must be able to describe the world in the same way (or in equivalent ways), disagreement between observers has been a main resource for theory discovery – contribution illustrated by (but not limited to) the formulation of Galilean and Lorentz transforms, and by all the physics that follows from them. In the same way, the requirement of physical salience has a crucial role in the evaluation of the status of surplus mathematical structures in physics, or in the identification of phenomena that require an explanation (identification and explanation of 'weird' or unexpected phenomena also acts as a pivotal heuristic guide in scientific discovery). These and other requirements had and keep having a major role in the evolution of physics, but it is not clear why should they play a role in a theory that, although investigates physical phenomena, does not describe physical reality, but only our state of belief.

It is legitimate therefore to wonder whether a theory that does not possess the same epistemic and methodological tools as physics, has the resources to produce physical knowledge and to solve physical problems rather than just making them disappear.

It has to be said that many advocates of the epistemic approach try to safeguard the objectivity of the theory by adding a physical dimension to their interpretation. For instance, advocates of QBism claim that, once all the 'epistemic' features will be distilled from the quantum state, the physical, quantum core will be found (Fuchs *et al.* 2014). However, this is not sufficient to meet our challenge. The appeal to such a physical core, in fact, does not play any role in the interpretation of the quantum state provided by these approaches, in their explanatory strategies, nor in their contents. It does not add any element that should be constrained by Koberinski's and Muller's criterion.

An unqualified appeal to a physical, objective, dimension in the theory is not sufficient as a solution to the challenge here posed. Such a challenge goes beyond the antirealism of the epistemic approach, but it hinges on the fact that their success is based on a *pass partout* that might potentially allow an arbitrary relaxation of the standards of scientificity for natural sciences. In order for an appeal to a physical content to meet our challenge, such a content must be shown to be constrained by general scientific criteria (like, again, testability, salience, or intersubjectivity) – it must show, in brief, that there is more to QT than formal epistemology.

In the same way, the problem persists when we move to epistemic but objective interpretations (e.g. Spekkens 2007), as far as the objective part does not make a substantial contribution in the theory, i.e. a contribution that will submit to the basic constraints of natural science. At the state of the art the challenge is not yet met by epistemic, subjective nor objective, approaches.

If we now get back to QT, the above considerations shed a different light on the solution to the problem of non-locality formulated by epistemic approaches. As the advocates of the epistemic view underlie (e.g. Fuchs *et al.* 2014), the necessity of explaining non-locality depends on the idea that QT describes physical reality: without said assumption, the problem of non-locality does not appear. In the literature devoted to antirealist approaches to QT, this disappearance is advertised as a success for these approaches. However, until when the above challenge is not met, it is legitimate to wonder whether such a result is a genuine success, or just the first of a series of dismissals of concrete physical problems that cannot be solved, but are hidden within an approach that does not have the means to formulate them.

**Acknowledgements**

The work of Jeff Bub has been highly influential on my work, and I am extremely thankful to him for the kindness and support that he has shown to me, while commenting this paper. He really is a great person. I am highly indebted with Michael Cuffaro for his thorough comments on this paper and extended discussions on this and related topics. I am also very thankful to Orly Shenker for her generous continuous support as for comments on an earlier version of this paper, and for the same reason to Mauro Dorato, Alexei Grinbaum and Matteo Morganti.